\documentstyle[12pt]{article}
\title{ Inclusive $\eta'$ production in B decays and the
Enhancement due to charged technipions }
\author{ Gongru Lu$^{a,b}$
\thanks{E-mail address: dphnu@public.zz.ha.cn },
Zhenjun Xiao$^{a,b,c}$, Hongkai Guo$^{b}$, Linxia L\"{u}$^b$ \\
\small a. CCAST(World Laboratory) P.O. Box 8730, Beijing 100080,
P.R.China \\
\small b. Department of Physics, Henan Normal University,
Xinxiang, 453002 P.R.China. \\
\small c. Department of Physics, Peking University, Beijing, 100871
 P.R.China. \\  }
\date{\today}

\begin{document}
\maketitle
\begin{abstract}
The new contributions to the charmless B decay $B \to X_{s}\eta^{\prime}$
from the unit-charged technipions $P^{\pm}$ and $P^{\pm}_{8}$ are estimated.
The technipions  can provide a large enhancement to the inclusive branching
ratio:  $Br(B \to X_{s}\eta^{\prime}) \sim 7\times 10^{-4}$ for
$m_{p1}=100GeV$ and $m_{p8}=250 \sim 350 GeV$ when the effect of QCD gluon
anomaly is also taken into account. The new physics effect is essential to
interpret the CLEO data.
\end{abstract}

\vspace{0.5cm}


\newcommand{\beq}{\begin{eqnarray}}
\newcommand{\eeq}{\end{eqnarray}}
\newcommand{\paa}{P^{\pm}}
\newcommand{\pbb}{P_8^{\pm}}
\newcommand{\betax}{Br(B \to \eta^{\prime} X_s)}
\newcommand{\betak}{Br(B^{\pm} \to \eta^{\prime} K^{\pm})}

\newpage

Recently CLEO has reported \cite{cleo98} a very large branching ratio
for the inclusive production of $\eta'$:
\beq
\betax =(6.2\pm 1.6\pm 1.3)\times 10^{-4},
\ \ for \ \  2.0\leq E_{\eta'}\leq 2.7GeV
\label{betax}
\eeq
and a corresponding large exclusive branching ratio \cite{cleo97}
\beq
\betak = (7.1 ^{+2.5}_{-2.1} \pm 0.9) \times 10^{-5}
\eeq
where the acceptance cut was used to reduce the background from events with
charmed mesons. By using the Standard Model (SM) factorization one finds
\cite{kagan97}
$\betax \sim (0.5 - 2.5) \times 10^{-4}$ including the experimental cut,
with the largest yields corresponding to a fairly limited region of
parameter space, which is much smaller than the observed inclusive rate in
eq.(\ref{betax}).

Up to now a number of interpretations have been proposed
\cite{datta98,yuan97,halperin97,atwood97,hou97,kagan97} to account for
the observed large branching ratio of $B \to \eta' X_s$ and/or
the exclusive branching fraction $\betak$.
These include: (a) conventional $b\to sq \bar q$ with constructive
interference
between the $u \bar u$, $d \bar d$ and $s\bar s$ components of the
$\eta^{\prime}$\cite{datta98}, (b) $b \to c\bar c s$ decay enhanced
$c\bar c$ content of the $\eta^{\prime}$ \cite{yuan97,halperin97}, (c)
$b\to s g^* \to s g \eta^{\prime}$ from QCD gluon anomaly \cite{atwood97}
or from both QCD gluon anomaly with running $\alpha_s$ and the new physics
effects \cite{hou97,kagan97}, (d) non-spectator effects\cite{du97}.

From the above works, the following major points about the inclusive and
exclusive branching ratios $\betax$ and $\betak$ can be reached;

\begin{enumerate}

\item
The SM factorization can, in principle, account for the exclusive
$\eta'$ yield without the need of new physics\cite{kagan97,du97}.
Although a SM "cocktail" solution for large inclusive rate $\betax$
involving contributions
from several mechanisms is still possible,  but the intervention of new
physics in the form of enhanced chromo-magnetic dipole operators provides
a simple and elegant solution to the puzzle in question\cite{hou97,kagan97}.
On the other hand, the short-distance $b \to \eta' s g$ subprocess
most possibly does not affect the exclusive $B \to \eta' K$ branching
ratios\cite{kagan97}.

\item
The observed inclusive branching fraction is larger than what is expected
from scenario (a). Furthermore, the data show that the
invariant mass spectrum $M(X_s)$ of the particles recoiling against the
$\eta'$ peaks above $2GeV$\cite{cleo98}.

\item
The large inclusive rate may be connected to the standard model QCD
penguins via the gluon anomaly, which leads to the subprocess $b \to sg^*
\to sg\eta^{\prime}$. Taking a constant $gg\eta'$ vertex form factor
$H(0,0,m_{\eta'}^{2})$ \cite{atwood97},
the observed large branching ratio in eq.(\ref{betax}) can be achieved. But
as argued by Hou and Tseng \cite{hou97}, if one considers the running of
$\alpha_{s}$ implicit in $H(0,0,m_{\eta'}^{2})$, the result presented
in ref.\cite{atwood97} will  be reduced by roughly a factor of 3. In
other words,  the new physics effect is essential to interpret the
observed large inclusive rate\cite{hou97}.

\item
As pointed by Kagan and Petrov \cite{kagan97},
the $m_{\eta'}^2/(q^{2}-m_{\eta'}^2)$ dependence of the
$gg\eta'$ coupling should be considered. Including  this dependence
nominally reduce the former result \cite{atwood97} to $\sim 1.6\times
10^{-5}$ including the cut, which is significantly smaller than the
observed inclusive rate. This fact will strengthen the need for new
physics.

\item
It is possible to enhance the chromo-dipole $bsg$ coupling by new physics at
the TeV scale without jeopardizing the electrodipole $bs\gamma$
coupling\cite{kagan95,ciuchini96}. One explicit example in the framework of
the Minimal Supersymmetric Standard Model (MSSM) has been studied
in ref.\cite{ciuchini96}.

\end{enumerate}

In this letter we will show that the unit-charged technipions $\paa$ and
$\pbb$ appeared in almost all nonminimal technicolor models
\cite{eichten86,lane96} can provide the required
enhancement to account for the observed large rate $\betax$\cite{cleo98}.

In the framework of the SM, the loop induced effective $bsg$ coupling
was calculated long time ago\cite{hou88},
\beq
\Gamma_{\mu}^{SM}= g_s\frac{G_F}{ 4\sqrt{2}\pi^2 } V_{is}^* V_{ib}
\overline{s} T^a \left[  F_1^i ( q^2 \gamma_{\mu} - q_{\mu} \not q )
-i \, F_2^i \sigma_{\mu\nu} q^{\nu}( m_s L + m_b R ) \right] b
\label{gasm}
\eeq
where the $g_s$ is the strong coupling constant, V is the CKM matrix,
$T^a=\lambda^a/2$ and $\lambda^a$ is the Gell-Mann matrix, $q=p_b -p_s$
and the charge radius form factors $F_1^i$ and the dipole moment $F_2^i$
( $i=u, c, t$ ) are
\beq
F_{1}^{i}&=&\frac{x_i}{12}\left[ y_{i}+13y_{i}^{2}-6y_{i}^{3}\right]
+\left[ \frac{2 y_i}{3}-\frac{x_i}{6}(4 y_{i}^{2}+
5 y_{i}^{3}-3 y_{i}^{4})\right] \ln[ x_{i}], \\
\label{f1sm}
F_{2}^{i}&=&-\frac{x_{i}}{4}\left[-y_{i}+3y_{i}^{2}+6y_{i}^{3}\right]
+\frac{3x_{i}^{2} y_i^3}{2}\ln [x_{i}]
\label{f2sm}
\eeq
where $x_{i}=m_{i}^{2}/M_{W}^{2}$ and $y_{i}=1/(x_{i}-1)$ for $i= u, c, t$.

In the framework of Technicolor theory, the new effective $bsg$ coupling
can be derived by replacing the internal W-lines in the one-loop diagrams
that induce $bsg$ coupling in the SM with the charged
technipion lines. Using the gauge and effective Yukawa couplings as given
in refs.\cite{eichten86}, one finds the new effective $bsg$ coupling induced
by $\paa$ and $\pbb$,
\beq
\Gamma_{\mu}^{New}=g_{s}\frac{G_F}{ 4\sqrt{2}\pi^{2} } V_{is}^* V_{ib}
\overline{s} T^a \left[ F_1^{(i, New)} ( q^2 \gamma_{\mu}
- q_{\mu}\not q )
- i \, F_2^{(i, New)} \sigma_{\mu\nu} q^{\nu}( m_s L + m_b R ) \right] b
\label{ganew}
\eeq
with
\beq
F_1^{New}(\xi_i, \eta_i)&=& \frac{D'(\xi_i)}{ 3\sqrt{2}G_F F_{\pi}^2 } +
\frac{8D'(\eta_i) }{ 3\sqrt{2}G_F F_{\pi}^2 } \\
F_2^{New}(\xi_i, \eta_i)&=&-\left[ \frac{D(\xi_i)}{3\sqrt{2}G_F F_{\pi}^2 }+
\frac{8D(\eta_i)+E(\eta_i)}{3\sqrt{2}G_FF_{\pi}^2} \right]
\eeq
and
\beq
D(\xi)&=&\frac{-5 + 19\xi-20\xi^2}{24(1-\xi)^3}
 + \frac{ 4\xi^2-2\xi^3 }{4(1-\xi)^4}\ln[\xi] \\
D'(\xi) &=& \frac{ 7- 29\xi + 16\xi^2 }{72(1-\xi)^{3}}-
\frac{3\xi^2- 2\xi^3}{12(1-\xi)^4}\ln[\xi] \\
E(\eta)&=&\frac{12-15\eta- 5\eta^2}{8(1-\eta)^3}
+\frac{ 9\eta-18\eta^2}{4(1-\eta)^4}\ln[\eta]
\eeq
where $\xi=m_{p1}^{2}/m_t^{2}$ and $\eta=m^2_{P8}/m_t^{2}$, and $m_{p1}$ and
$m_{p8}$ denote the masses of the color-singlet and color-octet technipion
$\paa$ and $\pbb$ respectively. The technipion decay constant $F_{\pi}=123
GeV$ in the One-Generation Technicolor Model (OGTM) \cite{eichten86}.
The $G_F$ is the Fermi coupling constant $G_F=1.16639\times10^{-5}
(GeV)^{-2}$.

Comparing the effective $bsg$ coupling  $\Gamma_{\mu}^{New}$ in
eq.(\ref{ganew}) with the $\Gamma_{\mu}^{SM}$ in eq.(\ref{gasm}),
one can see that the form factors $F_1^{(i, New)}$ and $F_2^{(i, New)}$
are the
counterparts of the $F_1^i$ and $F_2^i$ in the SM. The new form factors
$F_1^{(i, New)}$ and $F_2^{(i, New)}$ describe the contributions
to the decay $b \to sg$ from the charged technipions $\paa$ and $\pbb$.

   In the numerical calculation we use the branching ratio formula for
$B\rightarrow \eta'+ X_s$ with gluon anomaly as given in ref.\cite{hou97},
\beq
\frac{d^{2} Br( b\to \eta'sg ) }{dx dy} =
0.2\left[ \frac{g_s(m_b)}{4\pi^{2}} \right]^2
\frac{a_g^2 m_b^2}{4} \left[ |\Delta F_{1}|^{2} c_{0}+
Re(\Delta F_{1}F_{2}^*)\frac{c_{1}}{y}+|F_{2}|^{2}
\frac{c_{2}}{y^{2}} \right] \label{ffd}
\eeq
where $0.2$ comes from $(V_{cb}^{2}G_{F}^{2}m_{b}^{2})/(192\pi^{3})
\simeq 0.2\Gamma_{B}$ via the standard trick of relating to $B_{s.l}$
(see ref.\cite{atwood97}). The factors $c_0, c_1$ and $c_2$ in
eq.(\ref{ffd}) are
\beq
c_0&=&\left[-2x^{2}y+(1-y)(y-x')(2x+y-x')\right]/2, \nonumber\\
c_1&=&-(1-y)(y-x')^{2}, \nonumber\\
c_2&=&\left[ 2x^{2}y^{2}-(1-y)(y-x')(2xy-y+x')\right]/2
\label{ci}
\eeq
where $ x=m^2/m_b^2$ with $m$ is the physical recoil mass
against the $\eta'$ mason,  and $y=q^{2}/m_{b}^{2}$ with $q=p_b -p_s$
and $ x'=m_{\eta'}^2/m_b^2$. The term  $\Delta F_{1}$ was
defined as $\Delta F_1 = F_1(x_t) - F_1(x_c)$. The factor
$a_g=\sqrt{N_f} \alpha_s(\mu)/(\pi f_{\eta'})$ is the effective anomaly
coupling $H(q^2, k^2, m_{\eta'}^2)$ as defined in ref.\cite{atwood97}
and $f_{\eta'}=131MeV$. For the running of $\alpha_s$, we use the two-loop
approximation as given for instance in ref.\cite{buras97}.

The Fig.1 shows the mass dependence of form factors in the SM and in
the OGTM. The dot-dashed line corresponds to the $\Delta F_1=-5.25$
in the SM, while the long dashed line shows the $\Delta F_1^{New} =
F_1^{New}(\xi_t, \eta_t) - F_1^{New}(\xi_c, \eta_c) \approx -4.6$ in
the OGTM, assuming $m_{p1}=100GeV$ and $m_{p8}=250 - 600 GeV$.
The short-dashed line is the $F_2=0.2$ in the SM for $m_t=180GeV$ and
$m_W=80.2GeV$, while the solid curve
is the $F_2^{New}$ in the OGTM, assuming
$m_{p1}=100GeV$ and $m_{p8}=250 - 600 GeV$. It is easy to see that
the size of $F_2^{New}$ can be much larger than the $F_2$ in the SM
for light color-octet technipion. Furthermore, the $\pbb$
dominates the total contribution to the $F_1^{New}$ and $F_2^{New}$.

Because we do not know the "correct" form of $gg\eta'$ vertex form
factor $H(q^2,k^2,m^2_{\eta'})$, we consider the following
two different cases respectively.

\noindent{\bf Case-1:}
 We consider the effect due to  the running of $\alpha_{s}$
\cite{hou97} as well as the new contribution from the  charged technipions.

After the inclusion of the running of $\alpha_s$ one finds
$\betax \approx 3.4\times 10^{-4}$ including the
cut, as shown in Fig.2 (the dot-dashed line).
The horizontal band in Fig.2 represents the CLEO data in eq.(\ref{betax}).
The long dashed curve corresponds to the total inclusive branching ratio
$\betax$ when the new physics effects are also included. Numerically,
$\betax = (48.9 - 5.7 )\times10^{-4}$ for  $m_{p1}=100GeV$ and
$m_{p8}=(250 - 600)GeV$. The theoretical prediction is now well consistent
with the CLEO data for $m_{p8} \geq 350 GeV$. The color-octet technipion
$\pbb$ dominates the total new contribution:  the increase due to the $\paa$
is only about $10\%$ at the level of the corresponding branching ratio.

\noindent{\bf Case-2: }
We consider the effect of the $m_{\eta'}^2/(q^2- m_{\eta'}^2)$ suppression
and the new physics contribution from $\paa$  and $\pbb$.

When the new suppression factor $m_{\eta'}^2/(q^2- m_{\eta'}^2)$ is taken
into account one finds $\betax=2.3\times 10^{-5}$ including the cut as shown
by the short dashed line in Fig.2, which is much smaller than the CLEO
measurement. When the new contributions from the charged
technipions are included, the inclusive branching ratio $\betax$ can be
enhanced greatly as illustrated by  the solid curve in Fig.2. Numerically,
$\betax = (15.2 - 0.7)\times 10^{-4}$ for  $m_{p1}=100GeV$ and $m_{p8}=
(250 - 600)GeV$. The theoretical prediction is now  consistent
with the CLEO data for $m_{p8} \sim 280 GeV$. The new physics effect is
essential to interpret the CLEO data for the Case-2. Again, the color-octet
technipion $\pbb$ dominates the total contribution as that in Case-1.

In this letter we show a real example that the observed large ratio $\betax$
can be explained by the new physics contributions from the unit-charged
technipions
$\paa$ and $\pbb$. Because the major properties of the technipions in
different technicolor models are generally very similar, the analytical and
numerical results obtained in this letter are  representative and  can be
extended to other new technicolor models easily.

In this letter, we firstly evaluate the new one-loop penguin diagrams
with the internal $\paa$ and $\pbb$ lines and obtained the new form factors
$F_1^{New}(\xi_i, \eta_i)$ and $F_2^{New}(\xi_i, \eta_i)$ which describe the
new physics contributions to the decay in question. The size of $F_2^{New}$
can be rather large for relatively light charged technipions. Secondly, we
combine the new form factors $F_i^{New}$ (i =1,2) with  their counterpart
$F_1$ and $F_2$ in the SM properly and use them in the numerical calculation.
We finally calculate the inclusive branching
ratios for both Case-1 and Case-2. As illustrated in Fig.2, the  unit-charged
technipions  can provide a large enhancement to account for the large rate
$\betax$ observed by CLEO\cite{cleo98}.

\vspace{.5cm}

The authors would like to thank Dongsheng Du, Kuangta Chao and Yadong Yang
for helpful discussions. This work is supported  by the National Natural
Science Foundation of China under Grant No.19575015 and by the funds from
Henan Scientific Committee.

\vspace{1cm}

\vspace{1cm}
\begin{center}
{\bf Figure Captions}
\end{center}
\begin{description}

\item[Fig.1:] The dot-dashed line shows  $\Delta F_1=-5.25$
in the SM, the long dashed line corresponds to the $\Delta F_1^{New}$,
the short dashed line is the $F_2^{SM}=0.2$ and the solid curve
is the $F_2^{New}$ induced by unit-charged technipions.

\item[Fig.2:] The horizontal solid band represents the CLEO data $\betax =
(6.3 \pm 2.1)\times 10^{-4}$. The dot-dashed (short dashed) line is the
SM prediction in Case-1 (Case-2), while the long-dashed and solid line
show the total inclusive branching ratio $\betax$ when the new physics
effects are included in Case-1 and Case-2, respectively.

\end{description}

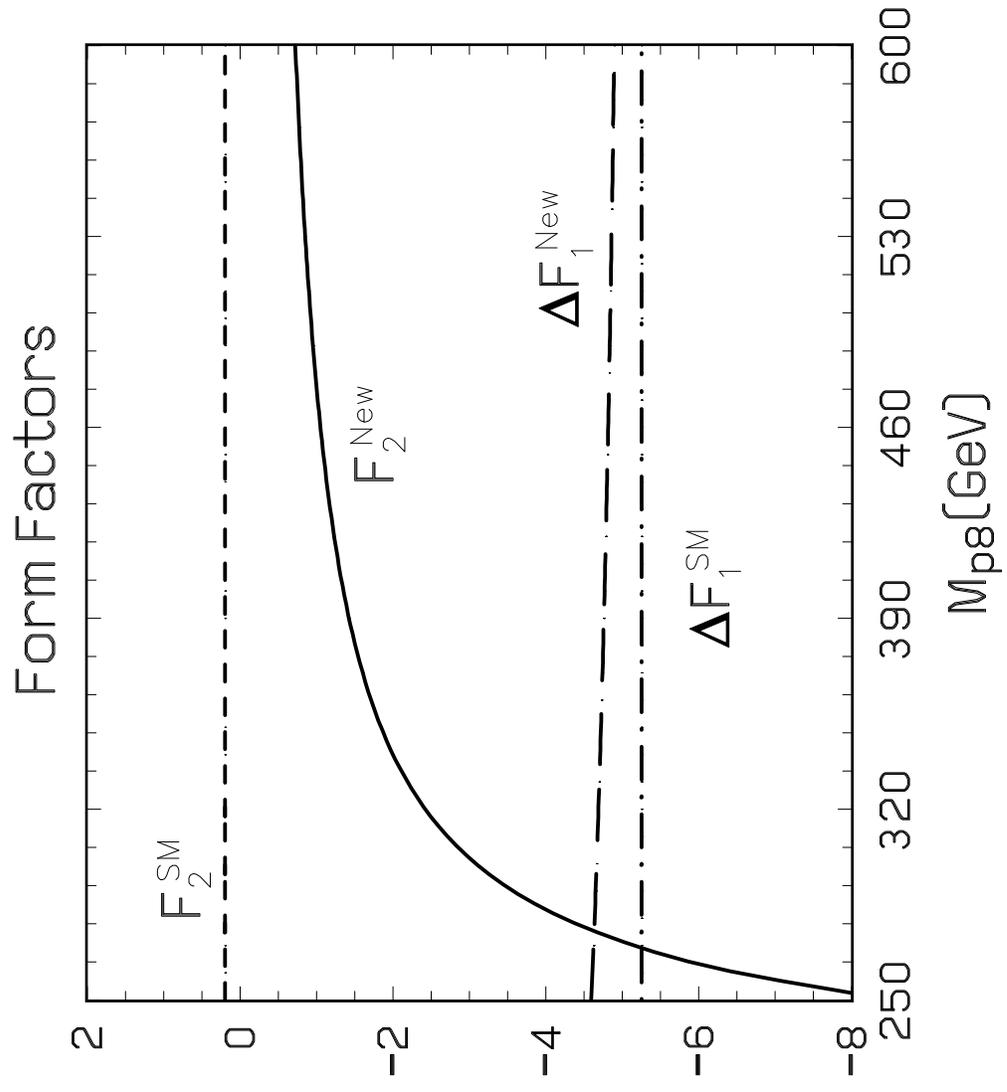
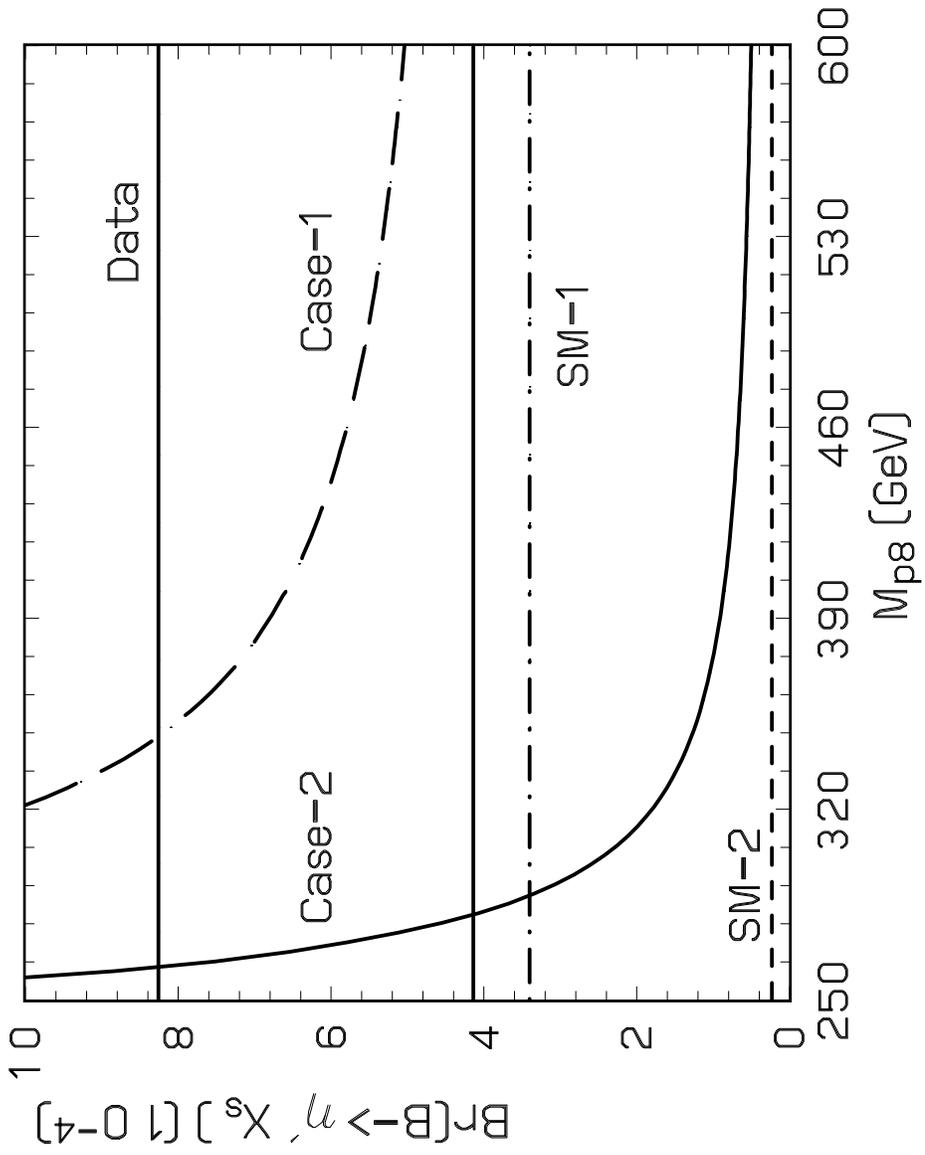
\end{document}